\documentclass[fleqn,twoside,twocolumn,nofootinbib]{revtex4} 
\usepackage[sec]{ujp} 

\begin{document}
\title[ANALOG OF THE COHERENT POPULATION TRAPPING STATE]
{ANALOG OF THE COHERENT POPULATION TRAPPING STATE IN A POLYCHROMATIC FIELD}%
\author{V.I. Romanenko}
\affiliation{Institute of Physics, Nat. Acad. of Sci. of
Ukraine}
\address{46, Prosp. Nauky, Kyiv 03028, Ukraine}
\email{vr@iop.kiev.ua}
\author{A.V. Romanenko}%
\affiliation{Taras Shevchenko National University of
Kyiv}%
\address{2, Prosp. Academician Glushkov, Kyiv 03022,
Ukraine}%
\author{L.P. Yatsenko}
\affiliation{Institute of Physics, Nat. Acad. of Sci. of
Ukraine}%
\address{46, Prosp. Nauky, Kyiv 03028, Ukraine}%
\email{vr@iop.kiev.ua}

 \udk{535.372} \pacs{42.50 Gy, 32.80 Wr}

\razd{\seciii}
\setcounter{page}{975}%
\maketitle


\begin{abstract}
The interaction between a three-level atom and a polychromatic field
with an equidistant spectrum ($\Lambda$-scheme of the atom--field
interaction) has been studied theoretically. It is shown that the
interaction of an atom with such a field can be reduced to its
interaction with a bichromatic field with additional light shifts of
transition frequencies and an additional coupling of the lower
atomic levels, which is proportional to the field intensity. Owing
to this coupling, the idea of the coherent population trapping can be
considered only as an approximation, because the dark state is not
an eigenstate of the effective Hamiltonian in the general case of
arbitrary dipole moments. The analyzed model gives a simple
theoretical interpretation for the formation of the atomic state, which
is close to the coherent population trapping, in the radiation field
of a femtosecond laser.
\end{abstract}

\section{Introduction}

\label{introduction}

The coherent population trapping (CPT) phenomenon
\cite{Ari76-333,Alz76-5,Gra78-218,Orr79-5,Alz79-209} is a basis for the
explanation of the electromagnetically induced transparency
\cite{Koc86,Gor89,Bol91-2593,Har97-36}, the population transfer among the
states by means of stimulated Raman adiabatic passage (STIRAP)
\cite{Ore84-690,Gau88-463,Ber98-1003,Vit01}, and the fabrication of compact
quantum laser-based frequency references \cite{Kna01-1545}. It also serves as
a basis for one of the methods aimed at the cooling of atoms down to ultralow
temperatures \cite{Asp88-826}.

In the simplest case, the CPT manifests itself at the interaction between a
three-level atom and a bichromatic radiation, when every spectral component
couples one of two metastable atomic states (one of them can be stable) with
the excited one ($\Lambda$-scheme of the atom--field interaction). Then, while
registering the dependence of the fluorescence intensity on the frequency of
either of spectral components, a narrow dip (a dark resonance) is observed,
when the difference between component frequencies becomes close to the
frequency of the transition between long-lived atomic states. From the physical
viewpoint, the CPT is based on the existence of a \textquotedblleft dark
state\textquotedblright\ in the atom, i.e. a superposition of long-lived
states, which is determined by the ratio between the intensities of spectral
components; if the atom is in this state, it does not absorb radiation.

Recently, the interest has been growing to dark resonances in a
polychromatic field with equidistant spectral components
\cite{Alz04,Sau,Ari06-169,Vla06-609,Auz}. The dark resonance arises
in the case where the frequency of the transition between long-lived
atomic states is a multiple of the frequency difference between the
neighbor spectral components of the field. Such resonances were
observed under the excitation of atoms with radiation emitted by a
femtosecond laser \cite{Auz}, whereas the resonance of the
electromagnetically induced transparency (which is based on the CPT
phenomenon) was registered in the field produced by a sequence of
light pulses generated by a mode locked picosecond laser. A dark
resonance with a contrast close to 100\% was also observed in sodium
vapor in the radiation field of a free-running multimode laser
\cite{Alz04}.

In this work, we analyze the interaction between a three-level
system and a low-intensity polychromatic field neglecting relaxation
processes and demonstrate that the polychromatic interaction can be
reduced to a bichromatic one, but with a certain field-dependent
frequency shift for transitions in the atom. In addition, we show
that, in the general case of arbitrary transition dipole moments,
one may talk only approximately about the CPT, because the dark
state is not an eigenstate of the effective Hamiltonian; namely, an
eigenstate of the effective Hamiltonian, which is the nearest to the
\textquotedblleft dark\textquotedblright\ one, almost always has a
small admixture of an excited state. The theoretical analysis of the
resonances in the CPT in a polychromatic laser field with a
frequency-shifted feedback \cite{Yat04-183} has already been carried
out by us taking advantage of the perturbation theory for the
density matrix and considering the uniform and non-uniform
broadening of spectral lines in the radiation emitted by atoms in a
buffer-gas environment \cite{Rom10-215402}. Although this research
allows one to obtain expressions for the frequency shifts of
transitions associated with a large number of spectral components,
it disregards, however, the physical essence of the CPT resonance,
namely, the formation of a non-absorbing or, more precisely,
low-absorbing state. To elucidate this issue, one has to find a form
for the effective Hamiltonian, which would describe very weak, but
numerous spectral components of the field.

\begin{figure}
\includegraphics[width=7.5cm]{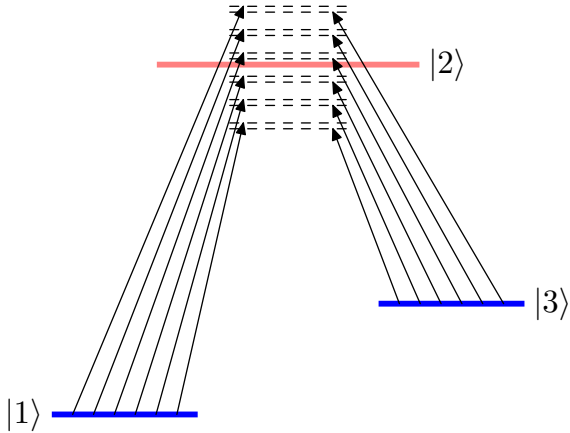}
\vskip-3mm\caption{Diagram of the interaction between a three-level
atom and a polychromatic field. The frequency difference between the
neighbor spectral components is equal to $\varpi$  }
\end{figure}

Our analysis is based on the averaging over the reciprocal frequency
difference between neighbor spectral components; it is the minimal
characteristic time $\tau$ of the problem. In other words, we
average over the quick motion of the system. The corresponding
formalism for a particle moving in a rapidly changing field is given
in work \cite{Landau}. Its modification to the case of a two-level
quantum-mechanical system was used to calculate the light pressure
force acting on the atom in a frequency-modulated field
\cite{Cas01}, as well as an atomic trap based on the interaction
between an atom and counter-propagating pulses \cite{Rom11-115305}.

In Section 2, the basic equations needed to describe the interaction
between an atom and the field are presented. Section~3 contains a
derivation of the effective Hamiltonian, which is responsible for a
slow -- with a characteristic time longer than $\tau$ -- variation
of probability amplitudes for the atom to remain in one of its
states. Light shifts of transition frequencies and the coupling
between metastable states, which is connected with the field
polychromaticity, are considered in Section~4 with the use of a
Gaussian-like distribution of the spectral intensity as an example.
In Section 5, one of the eigenstates of the effective Hamiltonian is
found, which is an analog of the CPT state in the case of
polychromatic field. In Section~6, CPT resonances in a polychromatic
field are studied making allowance for relaxation processes. The
results obtained are briefly summarized in Section 7.

\section{Basic Equations}

\label{basic-equations}

Consider a three-level atom with two metastable states, $|1\rangle$ and
$|3\rangle$, and one excited state $|2\rangle$, which interacts with the
polychromatic field characterized by equidistant frequencies between its
components (see Fig.~1). The frequency difference between neighbor spectral
components equals $\varpi$, and the frequency of the component with the maximum
intensity equals $\omega$.

The electric field $\bm{\mathcal{E}}(t)$ at the atom location point looks
like
\begin{equation}
\bm{\mathcal{E}}\left(  z,t\right)  =\frac{{\bm{\varepsilon}}\mathcal{E}_{0}%
}{2}\sum_{n=-\infty}^{\infty}a_{n}\exp\left[  -i\left(  \omega+n\varpi\right)
t-i\varphi_{n}\right]  +\text{c.c.}\label{eq:field}%
\end{equation}
Here, $\bm{\varepsilon}$ is the polarization unit vector, ${\mathcal{E}}_{0}$
the amplitude of the spectral component with the maximum intensity, and
$a_{n}$ are the relative amplitudes of spectral components. We normalize the
relative amplitudes $a_{n}$ in such a way that the component with $n=0$ has
the maximum amplitude, $a_{0}=1$. For model calculations, let us select a
realistic distribution of relative amplitudes in the form
\begin{equation}
a_{n}=\exp\left(  -\frac{n^{2}}{2n_{0}^{2}}\right)  .\label{eq:a}%
\end{equation}

In the rotating-wave approximation \cite{Sho90}, the interaction between the
three-level atom and the field is described by the Hamiltonian
\begin{equation}
\mathcal{H}=\frac{\hbar}{2}\left[
\begin{array}
[c]{ccc}%
-2\delta_{1} & \Omega_{1}^{\ast}(t) & 0\\
\Omega_{1}(t) & 0 & \Omega_{3}(t)\\
0 & \Omega_{3}^{\ast}(t) & -2\delta_{3}%
\end{array}
\right]. \label{eq:Ham}%
\end{equation}
Here, $\delta_{1}=\omega_{12}-\omega-n_{1}\varpi$ is a detuning of the
spectral component of laser radiation with the frequency $\omega+n_{1}\varpi$,
which is the nearest to the frequency $\omega_{12}$ of the transition
$|1\rangle\rightarrow|2\rangle$, from $\omega_{12}$, and $\delta
_{3}=\omega_{32}-\omega-n_{3}\varpi$ is a detuning of the spectral component
of laser radiation with the frequency $\omega+n_{3}\varpi$, which is the
nearest to the frequency $\omega_{32}$ of transition $|3\rangle\rightarrow
|2\rangle$, from the frequency $\omega_{32}$. The Rabi frequencies $\Omega
_{1}(t)$ and $\Omega_{3}(t)$ in Eq.~(\ref{eq:Ham}) are introduced by the
relations
\begin{equation}
\Omega_{j}(t)=\Omega_{j0}\sum_{n}a_{n}\exp\left[  i{}(n_{j}-n)\varpi
t-i\varphi_{n}\right]  ,\label{eq:Rabi-e}%
\end{equation}
where $j=1,3$, $\Omega_{j0}\equiv-(\bm{\varepsilon}\cdot\bm{d}_{j2}%
\,)\,\mathcal{E}_{0}/\hbar$, and $\bm{d}_{j2}$ is the matrix element of the dipole
moment, which is specifically selected to provide $\Omega_{j0}>0$.

\section{Effective Hamiltonian}

Let us suppose that
\begin{equation}
\Omega_{10},\;\Omega_{30},\;|\delta_{1}|,\;|\delta_{3}|\ll\varpi
.\label{eq:critD}%
\end{equation}
Hence, we consider the case of a so large frequency difference between the
neighbor spectral components of the field that the atom can be excited by only
two components with the relative amplitudes $a_{n_{1}}$ and $a_{n_{3}}$, the
frequencies of which are close to the transition frequencies $\omega_{12}$ and
$\omega_{32}$, respectively, whereas the other spectral component leads to a
light shift of resonance frequencies. The evolution of the wave function is
characterized by two time scales. The quick motion occurs with the
characteristic time $\varpi^{-1}$, and the slow one is determined by low
characteristic frequencies of the problem (see Eq.~(\ref{eq:critD})). To
obtain an equation describing the slow variation of probability amplitudes,
let us use the standard averaging procedure described in work \cite{Landau}
for the case of the mechanical motion in a field with a rapidly oscillating force.

Let us write down the Rabi frequencies (\ref{eq:Rabi-e}) in the form of sums of
two terms; one of those terms is time-independent, the other oscillates
with the frequency $\varpi$,
\begin{equation}
\Omega_{j}(t)=\Omega_{j,s}(t)+\Omega_{j,f},\qquad{}j=1,3,\label{eq:Rabis}%
\end{equation}
where
\begin{equation}
\Omega_{j,s}
(t)=\Omega_{j0}a_{n_{j}}e^{-i\varphi_{n_{j}}},\label{eq:Rabi-res}
\end{equation}
\begin{equation}
\Omega_{j,f} (t)= \Omega_{j0}\sum_{n\ne{}0}a_{n+n_{j}}\exp
\left[-i{}n\varpi t-i\varphi_{n+n_{j}} \right].
\end{equation}
The probability amplitudes for atomic states can expectedly be written down as
a sum of two terms,
\begin{equation}
C_{i}=\tilde{C}_{i}+c_{i},\qquad{}i=1,2,3,\label{eq:C}%
\end{equation}
the second of which is small in comparison with the first one and oscillates.
From the Schr\"{o}dinger equation
\begin{equation}
i\hbar\frac{d}{dt}\left(
\begin{array}
[c]{c}%
C_{1}\\
C_{2}\\
C_{3}%
\end{array}
\right)  =\mathcal{H}\left(
\begin{array}
[c]{c}%
C_{1}\\
C_{2}\\
C_{3}%
\end{array}
\right)
\end{equation}
we obtain
\[
i\dot{\tilde{C}}_{1}+\underline{i\dot{c}_{1}}=-\delta_{1}{\tilde{C}}_{1}-
\underline{\delta_{1}{{c}}_{1}}+
\frac{1}{2}\Omega_{1,s}^{*}\tilde{C}_{2}+
\]
\begin{equation}
+\underline{\underline{\frac{1}{2}\Omega_{1,f}^{*}c_{2}}
+\frac{1}{2}\Omega_{1,f}^{*}\tilde{C}_{2}}+\underline{\frac{1}{2}
\Omega_{1,s}^{*}c_{2}},\label{eq:Schi}
\end{equation}
\[
i\dot{\tilde{C}}_{2}+\underline{i\dot{c}_{2}}=
\frac{1}{2}\Omega_{1,s}\tilde{C}_{1}+\frac{1}{2}\Omega_{3,s}\tilde{C}_{3}+
\]
\[
+\underline{\underline{\frac{1}{2}\Omega_{1,f}c_{1}}
+\frac{1}{2}\Omega_{1,f}\tilde{C}_{1}+\frac{1}{2}\Omega_{1,s}c_{1}}+\label{eq:Schii}
\]
\begin{equation}
+\underline{\underline{\frac{1}{2}\Omega_{3,f}c_{3}}
+\frac{1}{2}\Omega_{3,f}\tilde{C}_{3}+\frac{1}{2}\Omega_{3,s}c_{3}},
\end{equation}
\[
i\dot{\tilde{C}}_{3}+\underline{i\dot{c}_{3}}=
-\delta_{3}{\tilde{C}}_{3}- \underline{\delta_{3}{{c}}_{3}}+
\frac{1}{2}\Omega_{3,s}^{*}\tilde{C}_{2}+
\]
\begin{equation}
+\underline{\underline{\frac{1}{2}\Omega_{3,f}^{*}c_{2}}
+\frac{1}{2}\Omega_{3,f}^{*}\tilde{C}_{2}}+\underline{\frac{1}{2}\Omega_{3,s}^{*}c_{2}}.\label{eq:Schiii}
\end{equation}
The terms underlined once rapidly oscillate and give zero after their
averaging over the oscillation period $2\pi/\varpi$. The terms underlined
twice consist of products of oscillating factors. Each of them can be divided
into a part that slowly varies in time and an oscillating part with a zero
averaged value. To exclude the rapid motion, let us postulate that the
corresponding terms on the left- and right-hand sides of Eqs.~(\ref{eq:Schi}%
)--(\ref{eq:Schiii}), which belong to the same type of the temporal
dependences, i.e. the rapid- or the slow-varying one, can be equated to each
other independently \cite{Landau}. The oscillating parts of
Eqs.~(\ref{eq:Schi})--(\ref{eq:Schiii}) bring about
\begin{equation}
i\dot{c}_{1}=
\frac{1}{2}\Omega_{1,f}^{*}\tilde{C}_{2},\label{eq:Schosci}
\end{equation}
\begin{equation}
i\dot{c}_{2}=\frac{1}{2}\Omega_{1,f}\tilde{C}_{1}+\frac{1}{2}\Omega_{3,f}
\tilde{C}_{3},\label{eq:Schoscii}
\end{equation}
\begin{equation}
i\dot{c}_{3}=\frac{1}{2}\Omega_{3,f}^{*}\tilde{C}_{2}.\label{eq:Schosciii}
\end{equation}
Here, we took inequalities $|c_{i}|\ll1$ and $|\dot{c}_{i}|\sim\varpi$, as
well as inequalities (\ref{eq:critD}), into account and neglected small terms.

The parts of Eqs.~(\ref{eq:Schi})--(\ref{eq:Schiii}), which are slowly varying in time,
give
\begin{equation}
i\dot{\tilde{C}}_{1}=-\delta_{1}\tilde{C}_{1}+
\frac{1}{2}\Omega_{1,s}^{*}\tilde{C}_{2}+
\frac{1}{2}\langle{\Omega_{1,f}^{*}c_{2}}\rangle, \label{eq:Schsli}
\end{equation}
\begin{equation}
i\dot{\tilde{C}}_{2}=\frac{1}{2}\Omega_{1,s}\tilde{C}_{1}+
\frac{1}{2}\langle{\Omega_{1,f}c_{1}}\rangle+
\frac{1}{2}\Omega_{3,s}\tilde{C}_{3}+\frac{1}{2}\langle{\Omega_{3,f}c_{3}}\rangle,
\label{eq:Schslii}
\end{equation}
\begin{equation}
i\dot{\tilde{C}}_{3}=-\delta_{3}\tilde{C}_{3}+
\frac{1}{2}\Omega_{3,s}^{*}\tilde{C}_{2}+
\frac{1}{2}\langle{\Omega_{3,f}^{*}c_{2}}\rangle,
\label{eq:Schsliii}
\end{equation}
where $\langle\cdots\rangle$ means time averaging over the interval
$2\pi/\varpi$.

The solution of Eqs.~(\ref{eq:Schosci})--(\ref{eq:Schosciii}) looks like
\begin{equation}
{c}_{1}=-\tilde{C}_{2}\Omega_{10}\sum\limits_{n\ne{}0}\frac{a_{n+n_{1}}}{2n\varpi}
\exp \left[i{}n\varpi
t+i\varphi_{n+n_{1}}\right],\label{eq:Schoscis}
\end{equation}
\[
{c}_{2}=\tilde{C}_{1}\Omega_{10}\sum\limits_{n\ne{}0}\frac{a_{n+n_{1}}}{2n\varpi}
\exp \left[-i{}n\varpi t-i\varphi_{n+n_{1}}\right]+
\]
\begin{equation}
+\tilde{C}_{3}\Omega_{30}\sum\limits_{n\ne{}0}\frac{a_{n+n_{3}}}{2n\varpi}
\exp \left[-i{}n\varpi
t-i\varphi_{n+n_{3}}\right],\label{eq:Schosciis}
\end{equation}
\begin{equation}
{c}_{3}=-\tilde{C}_{2}\Omega_{30}\sum\limits_{n\ne{}0}\frac{a_{n+n_{3}}}{2n\varpi}
\exp \left[i{}n\varpi
t+i\varphi_{n+n_{3}}\right],\label{eq:Schosciiis}
\end{equation}

Substituting the quantities ${c}_{i}$ into Eqs.~(\ref{eq:Schsli}%
)--(\ref{eq:Schsliii}), we arrive at equations for slowly varying amplitudes,
which have the form of a Schr\"{o}dinger equation with the effective
Hamiltonian
\begin{equation}
\mathcal{H}_{eff}=\frac{\hbar}{2}\left[
\begin{array}
[c]{ccc}%
-2\delta_{1}+S_{1} & \Omega_{1,s}^{\ast} & R\\
\Omega_{1,s} & -S_{1}-S_{3} & \Omega_{3,s}\\
R^{\ast} & \Omega_{3,s}^{\ast} & -2\delta_{3}+S_{3}\\
&  &
\end{array}
\right]  .\label{eq:hameff}%
\end{equation}
Here, the expressions
\begin{equation}
S_{1}=\frac{\Omega_{10}^{2}}{2\varpi}\sum_{n\neq{}0}\frac{|{a_{n+n_{1}}}
| ^{2}}{n}, \label{eq:eff-lsi}
\end{equation}
\begin{equation}
S_{3}=\frac{\Omega_{30}^{2}}{2\varpi}\sum_{n\neq{}0}\frac{|{a_{n+n_{3}}}
| ^{2}}{n}\label{eq:eff-lsiii}
\end{equation}
are responsible for light frequency shifts of the transitions $|1\rangle
\rightarrow|2\rangle$ and $|3\rangle\rightarrow|2\rangle$. These shifts are
determined by the spectral components of laser radiation, which are different
from $n_{1}$ or $n_{3}$ for the transition $|1\rangle\rightarrow|2\rangle$ or
$|3\rangle\rightarrow|2\rangle$, respectively, and the coupling between states
$|1\rangle$ and $|3\rangle$ determined by the coefficient
\begin{equation}
R=\frac{\Omega_{10}\Omega_{30}}{2\varpi}\sum_{n\neq{}0}\frac{a_{n+n_{1}%
}a_{n+n_{3}}}{n}\exp\left(  i{}\varphi_{n+n_{1}}-i{}\varphi_{n+n_{3}}\right)
.\label{eq:eff-R}%
\end{equation}
The latter has the dimensionality of frequency. The condition for the two-photon
resonance is obeyed, if
\begin{equation}
\delta_{3}=\delta_{1}+\frac{1}{2}(S_{3}-S_{1}).\label{eq:two-phot-res}%
\end{equation}

\section{Light Shifts and the Coupling Between Metastable States}

The calculations of the constants $S_{1}$, $S_{3}$, and $R$ will be carried
out for the Gaussian-like field spectrum (Eq.~(\ref{eq:a})) with a large
number of spectral components. In this case, let $n_{1}$ and ${n_{3}}$ be also
much smaller than $n_{0}$: $n_{1}\ll n_{0}$ and $n_{3}\ll n_{0}$. Then,
\[
a_{n+n_{j}}=\exp\left(-\frac{n^{2}}{2n_{0}^{2}}\right)
\biggl[1-\frac{nn_{j}}{n_{0}^{2}}-\frac{n_{j}^{2}}{2n_{0}^{2}}+
\]
\begin{equation}
+\frac{n_{j}^{2}n(n+n_{j})}{2n_{0}^{4}}-\frac{n^{3}n_{j}^{3}}{6n_{0}^{6}}\biggr]+\ldots,\label{eq:ani}
\end{equation}
where $j=1,3$ and
\begin{equation}
S_{j}=-\frac{n_{j}\sqrt{\pi}\Omega_{j0}^{2}}{n_{0}\varpi}\left(
1-\frac{2n_{j}^{2}}{3n_{0}^{2}}\right). \label{eq:S}%
\end{equation}
It is evident that $n_{3}$ and $n_{1}$ differ from each other by an integer
number $N$,
\begin{equation}
n_{3}=n_{1}-N, \label{eq:ndiff}%
\end{equation}
which is determined by the ratio between the quantities $\omega_{12}%
-\omega_{32}=\omega_{13}$ and $\varpi$. Let us introduce the two-photon
detuning
\begin{equation}
\delta=\omega_{13}-N\varpi=\delta_{1}-\delta_{3}, \label{eq:two}%
\end{equation}
where the integer $N$ is selected to minimize $|\delta|$. Then, the quantity
\begin{equation}
\delta=\frac{1}{2}\left(  S_{1}-S_{3}\right)
\end{equation}
demonstrates how much the CPT resonance is detuned from the accurate
two-photon one. This shift is minimal, if
\begin{equation}
n_{1}=\left[  \frac{\Omega_{30}^{2}N}{\Omega_{30}^{2}-\Omega_{10}^{2}}\left(
1-\frac{2\Omega_{10}^{2}N^{2}\left(  \Omega_{10}^{2}+\Omega_{30}^{2}\right)
}{3n_{0}^{2}\left(  \Omega_{30}^{2}-\Omega_{10}^{2}\right)  ^{2}}\right)
\right]. \label{eq:ni}%
\end{equation}
In this case,
\begin{equation}
n_{3}=\left[  \frac{\Omega_{10}^{2}N}{\Omega_{30}^{2}-\Omega_{10}^{2}}\left(
1-\frac{2\Omega_{30}^{2}N^{2}\left(  \Omega_{10}^{2}+\Omega_{30}^{2}\right)
}{3n_{0}^{2}\left(  \Omega_{30}^{2}-\Omega_{10}^{2}\right)  ^{2}}\right)
\right]  . \label{eq:niii}%
\end{equation}
Here, the square brackets mean the integer part of the expression, because $n_{1}$
is an integer number. In the case where the matrix elements of transition
dipole moments are identical, i.e., $\Omega_{30}^{2}=\Omega_{10}^{2}$, the
light shift of the two-photon resonance is described by the expression
\begin{equation}
\delta_{3}-\delta_{1}=\frac{N\sqrt{\pi}\Omega_{10}^{2}}{2n_{0}^{2}\varpi}.
\label{eq:eq_d}%
\end{equation}
For $\varphi_{n}=0$, we find, using Eqs.~(\ref{eq:eff-R}) and (\ref{eq:ani}),
that
\begin{equation}
R=\frac{\Omega_{10}\Omega_{30}(n_{1}+n_{3})}{2n_{0}\varpi}. \label{eq:R-e}%
\end{equation}
Comparing Eqs.~(\ref{eq:S}) and\ (\ref{eq:R-e}), we see that the quantities
$S_{1}$, $S_{3}$, and $R\ $are of the same order of magnitude.

Since $n_{1}+n_{3}\ll n_{0}$ and $\Omega_{10},\Omega_{30}\ll\varpi$, the
coupling between metastable states is very weak in comparison with that
between the metastable states, on the one hand, and the excited one, on the
other hand. Under the condition $n_{1}=-n_{3}$, i.e. when the frequency of the
spectral component with the maximum intensity is equal to $\omega=\frac{1}%
{2}(\omega_{12}+\omega_{32})$, the coupling between metastable states is absent.

\section{Analog of Coherent Population Trapping State in a Polychromatic
Field}

It is evident from the effective Hamiltonian (\ref{eq:hameff}) that,
in the general case, states $|1\rangle$ and $|3\rangle$ are coupled
with each other by means of the field. If $R=0$, the effective
Hamiltonian (\ref{eq:hameff}) is identical to the Hamiltonian of an
atom in a bichromatic field. In this case, if the two-photon
resonance is realized, one of the Hamiltonian eigenstates -- namely,
the state corresponding to the zero eigenvalue -- does not include
the excited state $|2\rangle$\ \cite{Ber98-1003,Vit01}. In the
previous section, it was shown that the quantity $R$ is small in
comparison with the Rabi frequencies. Therefore, in the case of the
two-photon resonance, one should expect that the population of the
excited states belonging to one of the eigenstates of the effective
Hamiltonian (\ref{eq:hameff}) would be low. Let us determine this
eigenstate. Let $\delta_{1}=S_{1}/2$, $\delta_{3}=S_{3}/2$, and let
the condition $\varphi_{n}=0$ be satisfied. Standard calculations
give the following equation for the characteristic $\lambda$-values
of the Hamiltonian:
\[
2\lambda(2\lambda-\Omega_{0})(2\lambda+\Omega_{0})+4\left(S_{1}+S_{3}\right)
\lambda^{2}-
\]
\begin{equation}
-\Omega_{0}^{2}R\sin2\theta-R^{2}\left(S_{1}+S_{3}+2\lambda\right)=0.
\label{eq:lambda}
\end{equation}
Here, we introduced the notations
\begin{equation}
\theta=\arctan\frac{\Omega_{1,s}}{\Omega_{3,s}},\qquad\Omega_{0}=\sqrt
{\Omega_{1,s}^{2}+\Omega_{3,s}}.\label{eq:theta}
\end{equation}
From Eq.~(\ref{eq:lambda}), it follows in the zeroth-order approximation in
$R$ and $S_{1}+S_{3}$ that
\begin{equation}
\lambda\approx 0, \label{eq:lambda-0}
\end{equation}
\begin{equation}
\lambda_{\pm}\approx \pm\frac{1}{2}\Omega_{0}.
\end{equation}
The characteristic value (\ref{eq:lambda-0}) corresponds to the state of
coherent population trapping. Let us analyze it in more details.
Consider the corresponding eigenstate. As the first approximation,
using Eq.~(\ref{eq:lambda}), we find
\begin{equation}
\lambda=-R\sin\theta\cos\theta=-R\frac{\Omega_{1,s}\Omega_{3,s}}{\Omega
_{1,s}^{2}+\Omega_{3,s}^{2}}.\label{eq:lambda-1}%
\end{equation}
The second approximation adds nothing new to expression (\ref{eq:lambda-1}).
Although it is easy to obtain the third approximation, we do not present it,
because it is not essential for the results obtained here. The eigenstate of
Hamiltonian, which corresponds to the characteristic value (\ref{eq:lambda-1}),
looks like
\begin{equation}
\psi=\frac{\Omega_{0}}{\sqrt{\Omega_{0}^{2}+R^{2}\cos^{2}2\theta}}\left(
\begin{array}
[c]{c}%
-\cos\theta\\
\displaystyle\frac{R}{\Omega_{0}}\cos2\theta\\
\sin\theta
\end{array}
\right)  .\label{eq:eigv}%
\end{equation}
It is evident that the population of the excited Hamiltonian eigenstate
(\ref{eq:eigv}) is low and amounts to $\left(  \Omega_{0}/\varpi\right)  ^{2}$
by the order of magnitude.

\section{Resonances of Coherent Population Trapping in a Polychromatic Field}

Proceeding from the effective Hamiltonian of an atom in a polychromatic field
described by Eq.~(\ref{eq:hameff}), one may expect, at first sight, that the
shift of the atomic fluorescence minimum with respect to the two-photon
resonance is determined by the difference between the light shifts of the
first and third atomic states. Unlike the ordinary formation of the CPT
resonance in a bichromatic field, when the latter couples states $1\rangle$,
$|2\rangle$ and $|3\rangle$, $|2\rangle$, Hamiltonian (\ref{eq:hameff})
demonstrates that the field also couples states $|1\rangle$, $|3\rangle$. Such
a coupling is similar to that between oscillators, which, as is known
\cite{Landau}, results in a shift of their characteristic frequencies.
Therefore, there is a need to analyze the consequences, which the presence
of the term $R$ in the effective Hamiltonian brings about.

For the analysis of the atomic fluorescence dependence on the two-photon
detuning, let us take advantage of the equation for the density matrix.
Interaction between the atom and the field is taken into account by
Hamiltonian (\ref{eq:hameff}), so that all we need is to include the influence
of relaxation processes on the density matrix evolution into consideration.
The application of the effective Hamiltonian formalism is eligible, if the
difference $\varpi$ between the neighbor spectral components of the
polychromatic field considerably exceeds the rate of relaxation processes.

Suppose that, owing to the spontaneous emission, the atom can transit
from excited state $|2\rangle$ into states $|1\rangle$ and $|3\rangle$ at the
rates $\gamma_{1}$ and $\gamma_{3}$, respectively. In addition, the relaxation
processes give rise to the relaxation of the coherence $\rho_{13}$ at a low rate
$\gamma_{0}$, which is much slower than $\gamma_{1}+\gamma_{3}$. As a result,
the equations for the density matrix look like
\[
 {\frac {d}{dt}} \rho_{{11}} =\frac{i}{2}\Omega_{{1}}
 \rho_{{12}} -\frac{i}{2}\Omega_{{1}} \rho_{{21}}
+\frac{i}{2}R \left(\rho_{{13}}- \rho_{{31}}\right)   +\gamma_{{1}}
\rho_{{22}},
\]
\[
 {\frac {d}{dt}}\rho_{{12}}  =\frac{i}{2}\left(2\delta_{{1}}-2S_{{1}} -S_{3}\right) \rho_{{12}}
+\frac{i}{2}\Omega_{{1}}\left(\rho_{{11}}-\rho_{{22}}\right) -
\]
\[
-\frac{i}{2}R\rho_{{32}}+\frac{i}{2}
\Omega_{{3}}\rho_{{13}}-\frac{1}{2} \left( \gamma_{{1}}+\gamma_{{3}}
\right) \rho_{{12}},
\]
\[
 {\frac {d}{dt}}\rho_{{13}} =\frac{i}{2}  \left( 2\delta_{{1}}
-2\delta_{3}+S_{3}-S_{{1}}\right) \rho_{{13}} +
\]
\[
+\frac{i}{2} R \left(\rho_{{11}} -\rho_{{33}}\right) + \frac{i}{2}
\Omega_{{3}} \rho_{{12}}-\frac{i}{2}\Omega_{{1}}\rho_{{23}}
-\gamma_{{0}}\rho_{{13},}
\]
\[
 {\frac {d}{dt}}\rho_{{23}}  =\frac{i}{2} \left( S_{{1}}+2S_{{3}} -2\delta_{3}\right)
 \rho_{{23}}-\frac{i}{2}\Omega_{{1}}\rho_{{13}} +
\]
\[
 +\frac{i}{2}\Omega_{{3}}\left(  \rho_{{22}} -\rho_{{33}}\right)
  -\frac{i}{2} R\rho_{{21}}-\frac{1}2{}\, \left( \gamma_{{1}}+\gamma_{{3}} \right) \rho_{{23}} ,
\]
\[
 {\frac {d}{dt}}\rho_{{33}}  =\frac{i}{2}\Omega_{{3}}\left(\rho_{{32}}
  -\rho_{{23}}\right)+\frac{i}{2}R\left(\rho_{{31}} -\rho_{{13}}\right)
+\gamma_{{3}}\rho_{{22}},
\]
\begin{equation}
\rho_{nm}=\rho_{mn}^{*},~~~ n,m=1,2,3,~~~
\rho_{11}+\rho_{22}+\rho_{33}=1.
 \label{eq:rho}
\end{equation}
To make the notations more compact, we did not mark the Rabi frequencies with
the identical additional subscript $s$.

To estimate the fluorescence signal, it is necessary to find an
expression for the population $\rho_{22}$ in excited state. The CPT
resonance width is small, provided that the Rabi frequencies
$\Omega_{{1}}$ and $\Omega_{{3}}$ are of the order of $\gamma_{0}$.\
As is seen from expressions (\ref{eq:S}) and (\ref{eq:R-e}), the
light shifts and the frequency $R$ describing the coupling between
states $|1\rangle$ and $|3\rangle$ are small in comparison with the
Rabi frequencies. The latter are low in comparison with $\gamma_{1}$
and $\gamma_{3}$. Therefore, we assume that
\[
R/\Omega_{j}  \ll1,\quad  S_{j}/\Omega_{j} \ll1, \quad \Omega_{j}
\ll \gamma_{1} +\gamma_{3},
\]
\begin{equation}
\Omega_{j}\sim \gamma_{0}, \quad j=1,3. \label{eq:eff-crit}
\end{equation}
Under those conditions, the shift of the population maximum  in the
excited state, which we find from the stationary solution of
Eqs.~(\ref{eq:rho}), is minimal, if
\begin{equation}
\delta=\delta_{R}+\delta_{S},\label{eq:dmin}%
\end{equation}
where
\[
\delta_{R}={\Omega_{1}\Omega_{3}R\gamma_{0}
\left(\Omega_{3}^4\gamma_{1}-\Omega_{1}^4\gamma_{3}\right)\left(\gamma_{1}+\gamma_{3}\right)}\times
\]
\[
\times\biggl[
\Omega_{1}^{2}\Omega_{3}^{2}\left(\Omega_{1}^{2}+\Omega_{3}^{2}\right)
\left(\Omega_{1}^{2}\gamma_{3}+\Omega_{3}^{2}\gamma_{1}\right)
\vphantom{\left(\right)^{2}}-
\]
\begin{equation}
-\left(\Omega_{1}^{2}-\Omega_{3}^{2}\right)\left(\Omega_{1}^{2}\gamma_{3}-\Omega_{3}^{2}
\gamma_{1}\right)\left(\gamma_{1}+\gamma_{3}\right)^{2}R^{2}\biggr]^{-1}.
\label{eq:resR}
\end{equation}
Provided that conditions (\ref{eq:eff-crit}) are satisfied for a wide enough
spectrum of laser radiation, when $\omega_{13}/n_{0}\varpi\ll\varpi
/(\gamma_{1}+\gamma_{3})$, the second term in the square brackets in
Eq.~(\ref{eq:resR}) is much less than the first one; then, by the order of
magnitude, $\delta_{R}\sim(\gamma_{1}+\gamma_{3})R/\gamma_{0}$. In expression
(\ref{eq:dmin}), the second term plays a dominating role, as a rule, in those
partial cases where the first term is small, e.g., if $R=0$. Then,
\begin{equation}
\delta_{S}=\frac{1}{2}\left(  S_{1}-S_{3}\right)  .\label{eq:resS}%
\end{equation}

The obtained expression for the resonance shift is valid in the case of a cell
without buffer gas, when the relaxation rate of atoms in the excited state is
governed by the spontaneous emission, and all relaxation rates are much
slower than $\varpi$. In the presence of a buffer gas, the relaxation rate for
the optical coherences considerably exceeds $\varpi$. Therefore, the interaction
between the atom and the polychromatic field cannot be characterized any more
by an effective Hamiltonian. Such a case was analyzed in work
\cite{Rom10-215402}.

If the radiation spectrum width considerably exceeds $\omega_{13}$,
the Rabi frequencies of spectral components, which are resonant to
the atomic transition frequency, differ very slightly from the
frequency of the Rabi component with the maximal intensity. Taking
into account that they are proportional to the dipole moment of the
corresponding transition and the probability of the transition from
the excited state into state $|1\rangle$ or $|3\rangle$ is
proportional to the square of this quantity, we easily obtain the
relation
\begin{equation}
\frac{\Omega_{1}^{2}}{\gamma_{1}}=\frac{\Omega_{3}^{2}}{\gamma_{3}%
},\label{eq:Rabialt}%
\end{equation}
so that Eq.~(\ref{eq:resR}) takes the form
\begin{equation}
\delta_{R}=\frac{\gamma_{3}-\gamma_{1}}{2\Omega_{1}\Omega_{3}}\gamma
_{0}R.\label{eq:resRalt}%
\end{equation}
Hence, $\delta_{R}=0$, if $\gamma_{3}=\gamma_{1}$. In this case, the resonance
shift (\ref{eq:dmin}) is determined by the term $\delta_{S}$.

In Fig. 2, the dependences of population $\rho_{22}$ in the excited state on
the two-photon detuning $\delta$ from the resonance are depicted. They were
obtained by solving Eqs.~(\ref{eq:rho}) for the density matrix under condition
(\ref{eq:Rabialt}). As is evident from the figure, the position of the excited
state population minimum at $R\neq0$ obtained from the solution of the
equations for the density matrix agrees well with the result of calculations by
formula (\ref{eq:resRalt}). The minimum of the curve ($\delta_{S}%
=-0.005\gamma_{0}$) obtained for the case $R=0$ also agrees well with the
result of calculations by formula (\ref{eq:resS}), which is valid in this case.

\begin{figure}
\includegraphics[width=\column]{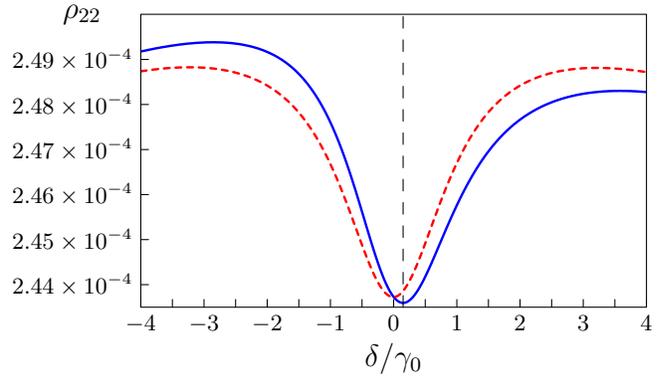}
\vskip-3mm\caption{Dependences of population $\rho_{22}$ in the
excited state on the two-photon detuning $\delta$ from the resonance
obtained from the solution of
Eqs.~(\ref{eq:rho}) for the density matrix under conditions (\ref{eq:Rabialt}%
). The calculation parameters are $\Omega_{1}=\gamma_{0}$, $\Omega_{3}%
=2\gamma_{0}$, $\gamma_{1}+\gamma_{3}=100\gamma_{0}$,
$S_{1}=0.01\gamma_{0}$, $S_{3}=0.02\gamma_{0}$, and $\delta_{1}=0$.
The solid curve corresponds to $R=0.01\gamma_{0}$, the dashed one to
$R=0$. The vertical line denotes the position of the excited state
population minimum ($\delta_{R}=0.15\gamma_{0}$) calculated by
formula (\ref{eq:resRalt})  }
\end{figure}

\section{Conclusions}

To summarize, we showed that, if a three-level atom interacts with a
polychromatic field with equidistant frequencies, there emerges a state, which
is similar to that of coherent population trapping in a bichromatic field. In
contrast to the case of bichromatic field, this state, besides metastable
states, also includes an insignificant admixture of the excited state, the
amplitude of which is smaller, the wider is the spectra of laser radiation.
The effective Hamiltonian, which describes a slow, in comparison with the
field period, variation of the probability amplitudes for atomic state
populations, contains a coupling between the metastable states. Such a
coupling gives rise, generally speaking, to a resonance shift in the coherent
population trapping.

An application of the developed theory to the creation of a frequency
reference on the basis of an ensemble of cold $^{229}$Th atoms or ions
\cite{Pei03} could be of interest. The corresponding split in the nuclear
ground state is about 7.6~eV \cite{Bec03} (vacuum ultraviolet), and the
lifetime of the nuclear excited state (an hour by the order of magnitude)
\cite{Tka00} testifies in favor of the creation of a high-precision clock on
the basis of the transition in this nucleus.

\vskip3mm The work was fulfilled as a part of the goal-oriented
themes V136 and VTs139 of the National Academy of Sciences of
Ukraine. It was also supported in the framework of the project
F~40.2/039 of the State Fund for Fundamental Researches.

\rezume{%
АНАЛОГ СТАНУ КОГЕРЕНТНОГО ПОЛОНЕННЯ\\ НАСЕЛЕНОСТЕЙ У
ПОЛІХРОМАТИЧНОМУ ПОЛІ}{В.І. Романенко, О.В. Романенко, Л.П. Яценко}
{Теоретично досліджено взаємодію трирівневого атома з
поліхроматичним полем з еквідістантними спектральними компонентами
($\Lambda$-схема взаємодії з полем). Показано, що взаємодія атома з
таким полем зводиться до взаємодії з біхроматичним полем з
додатковими світловими зсувами частот переходів і додатковим,
пропорційним інтенсивності поля, зв'язком нижніх рівнів атома.
Наявність цього зв'язку дозволяє говорити про когерентне полонення
населеності лише наближено, оскільки у загальному випадку довільних
дипольних моментів переходів темний стан не є власним станом
ефективного гамільтоніана. Розглянута модель дає просту теоретичну
інтерпретацію формування близького до когерентного полонення
населеностей стану атома у полі випромінювання фемтосекундного
лазера.}

\end{document}